\documentclass[twocolumn,showpacs,preprintnumbers,amsmath,amssymb]{revtex4}

\usepackage{graphicx}% Include figure files
\usepackage{dcolumn}% Align table columns on decimal point
\usepackage{amsfonts,amsmath,amssymb,bm}

\begin{document}

\title{Resonant magnetoresistance in organic spin-valves}

\author{A. Reily Rocha and S.~Sanvito}
\email{sanvitos@tcd.ie}
\affiliation{School of Physics, Trinity College, Dublin 2, Ireland}

\date{\today}

\begin{abstract}
We investigate theoretically the effects of surface states over the magnetoresistance 
of Ni-based organic spin-valves. In particular we perform {\it ab initio} electronic transport
calculations for a benzene-thiolate molecule chemically attached to a Ni [001] surface
and contacted either by Te to another Ni [001] surface, or terminated by a thiol group and
probed by a Ni STM tip. In the case
of S- and Te-bonded molecules we find a large asymmetry in the spin-currents as a function of
the bias, although the $I$-$V$ is rather symmetric. This leads to a smooth although not monotonic
dependence of the magnetoresistance over the bias. In contrast, in the case of a STM-type geometry 
we demonstrate that the spin-current and the magnetoresistance can be drastically changed
with bias. This is the result of a resonance between a spin-polarized surface state of
the substrate and the $d$-shell band-edge of the tip.
\end{abstract}


\maketitle

\section{Introduction} 
The electron spin has recently made its appearance in organic electronics, generating a new
field which takes the suggestive name of molecular-spintronics \cite{SS-Review}. The main idea
is to investigate how spin-currents propagate through organic molecules possibly affecting
their internal state. Example of such spin-molecular devices are spin-valves made from carbon
nanotubes \cite{bruce}, organic insulators \cite{Ralph} or conducting polymers \cite{Shi,Dediu},
fempto-second spectroscopy of molecular bridges connecting semiconductor nanodots \cite{Aws3} and 
three terminal devices using magnetic molecules as active elements \cite{mmtrans1,mmtrans2}.

Organic molecules present genuine advantages over conventional inorganic metals and 
semiconductors. Typically the spin-orbit interaction, which scales with the atomic number $Z$
as $Z^4$, is weak in molecules. For instance the spin-orbit splitting of 
the valence band of carbon in its diamond form is only 13~meV \cite{SOCarbon}, to be compared with
340~meV of GaAs. This yields rather long spin-diffusion
lengths ($l_\mathrm{s}$) with conservative estimates of 130~nm in multiwall carbon nanotubes \cite{bruce}, 
200~nm in polymers \cite{Dediu}, and 5~nm in Alq$_3$ monolayers \cite{Alq3}. It is important to
note that these values are usually extracted from the magnetoresistance (MR) of spin-valves using the 
Jullier's formula \cite{Jullier} and therefore also include possible spin-flip at the surface of the 
magnetic electrodes. This means that the intrinsic spin-diffusion length is such organics may be 
considerably longer. Furthermore one should remember that at least in polymers and Alq$_3$
the resistivity is rather large and therefore the spin-relaxation times associated to the values of
$l_\mathrm{s}$ given above are quite long.

In organic molecules also the hyperfine interaction, another possible
source of spin-decoherence, is weak. Most of the molecules used for spin-transport are 
$\pi$-conjugate and the conduction is mostly through molecular states with amplitude
over the carbon atoms. Carbon, in its most abundant isotopic form, $^{12}$C,
has no nuclear spin, and therefore is not hyperfine active. Moreover the $\pi$-states are usually 
delocalized and the interaction is in general weak even over those nuclei with non-vanishing
nuclear spin.

In addition to the long spin-diffusion length organic materials offer a range of new physical 
properties, difficult to find and tune in conventional inorganic materials. For instance the
electrical conductivity of a polymer can be changed by over twelve orders of magnitude 
with the incorporation of various counterions \cite{SSH}. 
More interestingly the current-carrying elementary excitations are not band-like but usually 
are in the form of electronic-vibronic correlated states, with peculiar spin characteristics \cite{Bishop}.
Furthermore magnetic molecules with collective high-spin ground states are available \cite{roberta}.
In these the transport can be dominated by long-living metastable states, giving rise to 
spin-dependent non-linearities in the $I$-$V$ characteristics \cite{mmtrans1,mmtrans2}.

Finally a less explored avenue is that of exploiting the many alternatives for bonding a molecule
to a magnetic surface. The electronic structure of the anchoring group can in fact change
quite drastically the magneto-transport response of a molecule \cite{Smeagol} paving the way
for a chemical design of magnetic devices. This is precisely the problem we address in 
the present work. In particular we investigate how spin-polarized surface states, almost
always present in the case of organic spin-valves, can affect the transport. 

%%%%%%%%%%%%%%%%%%%%%%%%%%%%%%%%%%%%%%%%%%%%%
%%%%%%%%%%%%%%%%%%%%%%%%%%%%%%%%%%%%%%%%%%%%%

\section{Computational Method}
All the calculations presented in this work have been performed with the code 
{\it Smeagol} \cite{Smeagol,Smeagol1}. {\it Smeagol} combines the non-equilibrium 
Green's function (NEGF) method \cite{datta,Haug} with the Kohn-Sham \cite{KS} form of
density functional theory (DFT) \cite{DFT}. In its present implementation {\it Smeagol} uses 
the code SIESTA \cite{siesta} as the main DFT platform. This has the advantage of expanding
the Kohn-Sham problem over an efficient linear combination of localized numerical orbitals
\cite{MZeta,MZeta1}, making large systems tractable without the need of massive computational
resources. In addition core electrons are removed from the self-consistent calculation and 
described by standard norm-conserving scalar relativistic Troullier-Martins pseudopotentials.

The central quantity that {\it Smeagol} evaluates is the non-equilibrium Green's function $G(E)$ 
for the portion of the device where the potential drops. Usually this is called  ``extended molecule'' 
(EM) and comprises the molecule itself and part of the current/voltage probes. Such a Green's function
simply reads 
\begin{equation}
G(E)=\lim_{\eta\rightarrow 0}[(E+i\eta)-H_\mathrm{EM}-\Sigma_\mathrm{L}-\Sigma_\mathrm{R}]^{-1}\:,
\label{negfmx}
\end{equation}
where $H_\mathrm{EM}$ is the Hamiltonian of the EM, $E$ is the energy and $\Sigma_\mathrm{L}$
($\Sigma_\mathrm{R}$) is the self-energy for the left-hand side (right-hand side) electrode.
These latter ones are non-hermitian energy-dependent matrices describing the interaction of the EM 
with the current/voltage 
electrodes and effectively allow us to map a hermitian problem for an open non-periodic system
onto a non-hermitian problem for a finite system. They can be written as 
$\Sigma_{\alpha}=H^\dagger_{\alpha\mathrm{EM}}g_\mathrm{\alpha}H_{\alpha\mathrm{EM}}$
with $H_{\alpha\mathrm{EM}}$ the coupling matrix between the leads $\alpha$ ($\alpha=$L, R) and the
EM. The $g_\alpha$ are surface Green's functions for the leads, i.e. the Green's
function for a semi-infinite lead evaluated at the termination plane. They are constructed
using a semi-analytical expression \cite{rgf}, after removing the singularities of the coupling matrices
with a renormalization scheme \cite{Smeagol}.

Importantly when $H_\mathrm{EM}$ is a function of the charge density $\rho$, as in DFT,
then a simple prescription for a self-consistent calculation can be designed. In fact, a close formula
for the non-equilibrium charge density in terms of the Green's function $G(E)$ is available \cite{Smeagol}, 
and can be used in conjunction with equation (\ref{negfmx}) for deriving the electronic structure of the
EM in presence of the leads. External bias can be added in the self-consistent procedure by imposing
the appropriate boundary conditions to the Hartree potential in the leads. This simply produces a rigid
shift of the whole spectra in the current/voltage electrodes, since they are good metals and therefore local charge neutral. 

Finally the two probe current can be extracted by using a Landauer-like formula
\begin{equation}
I=\frac{e}{h}\int_{-\infty}^{+\infty}\mathrm{d}E\:\mathrm{Tr}[G\Gamma_\mathrm{L}
G^\dagger\Gamma_\mathrm{R}][f(E-\mu_\mathrm{L})-f(E-\mu_\mathrm{R})]\;,
\label{CurrNEGF}
\end{equation}
where $G=G(E)$, $\Gamma_\alpha=i[\Sigma_\alpha-\Sigma_\alpha^\dagger]$, $f_\alpha(x)$
is the Fermi function and $\mu_\mathrm{L/R}=E_\mathrm{F}\pm eV/2$ with $E_\mathrm{F}$
the Fermi energy of the leads and $V$ the external bias. Note that ultimately the current is
nothing but the integral of the bias-dependent transmission coefficient $T(E)=G\Gamma_\mathrm{L}
G^\dagger\Gamma_\mathrm{R}$ evaluated over the bias window.

In our simulations the EM includes three Ni planes at each side of the molecule and the cross
section is large enough to prevent intermolecular interaction. We use periodic boundary conditions
in the direction orthogonal to the transport with a 9 k-points in the two-dimensional Brillouin zone. 
The real-space grid cut-off is fixed at 450~Ry \cite{siesta} and multiple-$\zeta$ basis sets are
constructed in the following way: double-$\zeta$ for the $s$ shell of H, C, S, Te and Ni, and for the
Ni $d$ shell, double-$\zeta$ plus polarization for the $p$ shells of C, S and Te, and
single-$\zeta$ for Ni $p$ orbitals. Finally the charge density is obtained by integrating the 
Green's function over 96 imaginary and 300 real energy points \cite{Smeagol}.

%%%%%%%%%%%%%%%%%%%%%%%%%%%%%%%%%%%%%%%%%%%%%
%%%%%%%%%%%%%%%%%%%%%%%%%%%%%%%%%%%%%%%%%%%%%

\section{Resonant transport in organic spin-valves}

Consider a typical spin-valve constructed from a strong ferromagnet and a non-magnetic molecule. 
In this set-up the transport is determined by the alignment of the molecular levels with the Fermi energy
($E_\mathrm{F}$) of the electrodes \cite{datta,SS-Review}. 
Let us consider the situation in which the transmission through the 
molecular level is small, but the anchoring groups form a spin-polarized localized 
state with the $d$-orbitals of the magnetic electrodes. The spin-splitting of such state will be comparable 
with the band-exchange of the ferromagnet, although its actual form and position will depend on the
details of the anchoring structure. 
\begin{figure}[ht]
\includegraphics[width=0.45\textwidth]{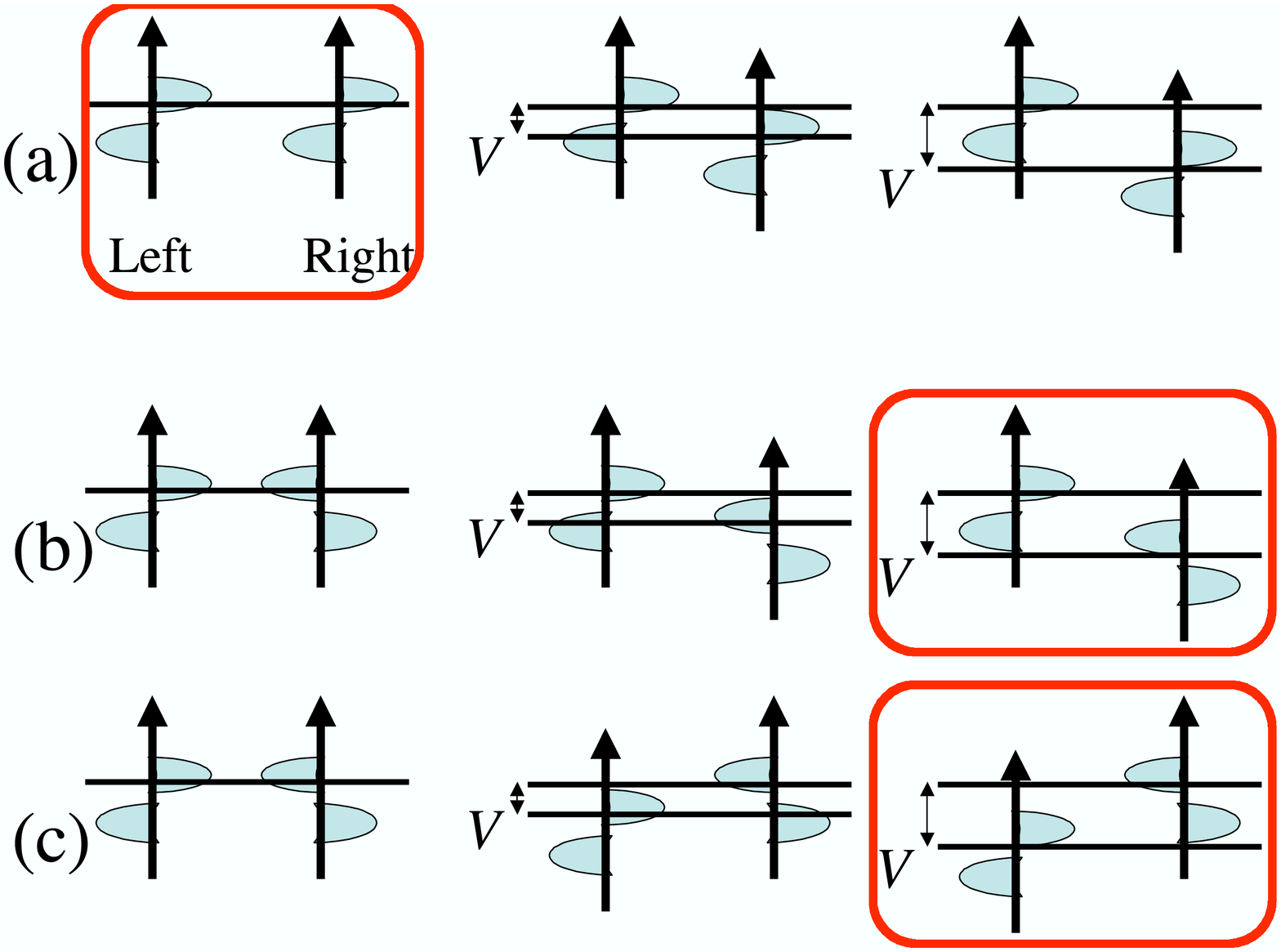}
\caption{\label{Fig1}
Cartoon representing the alignment of the spin-polarized surface states associated to the
two magnetic electrodes forming a spin-valve: (a) parallel alignment and forward bias,
(b) antiparallel alignment and forward bias, (c) antiparallel alignment and reverse bias. Panels
to the left are for $V$=0 and the bias increases from left to right. The red boxes mark the
three resonant conditions.}
\end{figure}

In this condition of strong coupling the position of such surface state (SS) is strongly pinned at
the chemical potential of the relevant electrode and shifts rigidly with the bias. Then, the 
alignment of the SS associated to the left electrode of the device with that of the right 
determines the current. In particular for identical electrodes and anchoring groups
there are three different resonant conditions (see figure \ref{Fig1}) depending
on the bias and the mutual alignment of the magnetization vectors in the contacts. In the case
of parallel alignment (PA) the state associated to the left-hand side electrode is lined up with
that of the right only at zero bias and for both spin directions (left panel of figure \ref{Fig1}a).
If it is also located at $E_\mathrm{F}$, then the current will be large around $V$=0. A finite bias 
reduces the overlap between the states, regardless of the bias polarity (in the picture we report
forward bias only). This means that the polarization of the currents decreases with bias 
and approximately vanishes for a bias voltage as large as the state width. 

The other two resonant conditions are found for the antiparallel alignment (AA) of the electrodes.
In fact the SSs on the two sides of the device overlap for a bias of the order of the exchange
splitting of the state itself. For forward bias such overlap is between the majority component of the
SS to the left with the minority of that to the right (last panel to the right of figure \ref{Fig1}b).
The opposite is for reverse bias (last panel to the right of figure \ref{Fig1}c).
This gives the interesting result that for the AA both the spin-currents
are highly asymmetric as a function of bias, although the total current is not. Thus, if we define 
the MR ratio as $R_\mathrm{MR}=(I_\mathrm{PA}-I_\mathrm{AA})/
I_\mathrm{AA}$, we will expect a minimum in $R_\mathrm{MR}$
for biases of the order of the SS exchange splitting. This occurs in correspondence of a large
current in the AA configuration.

A completely different situation is encountered when the attachment of the molecule
to a surface is highly asymmetric. In this case different resonance conditions will be 
present depending on the details of the structure and both the spin-current and the MR 
might have non-trivial bias dependence \cite{usPC}. Similar arguments have been also used 
for explaining rectification in non-magnetic molecular junctions using electrodes with
$d$ states close to $E_\mathrm{F}$ \cite{Kir1}.

%%%%%%%%%%%%%%%%%%%%%%%%%%%%%%%%%%%%%%%%%%%%%
%%%%%%%%%%%%%%%%%%%%%%%%%%%%%%%%%%%%%%%%%%%%%

\section{S-Benzene-T\lowercase{e} between N\lowercase{i} surfaces}

We start our analysis by investigating a benzene molecule chemically attached to the hollow 
site of two Ni [001] surfaces. As anchoring groups we use S on one side and Te on the other. 
These both form a stable chemical bond with Ni, although the two bond lengths are 
quite different (1.28~\AA\  for S and 1.77~\AA\ for Te) reflecting the rather different atomic radii of 
the anchoring atoms. With this configuration we expect an asymmetric $I$-$V$ characteristics,
although we do not expect any rectification, since the coupling with the electrodes is
strong.

\begin{figure}[ht]
\includegraphics[width=0.45\textwidth]{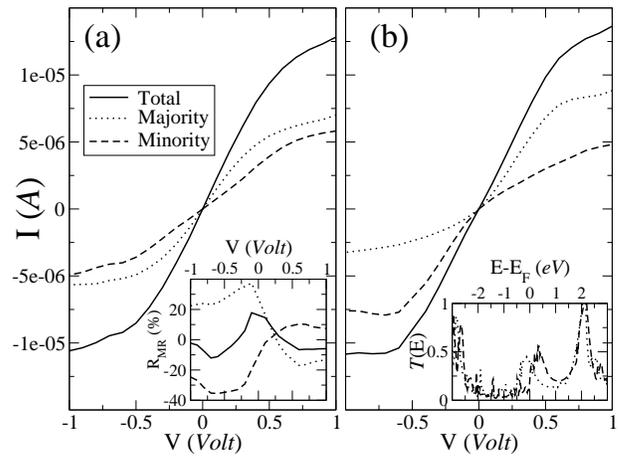}
\caption{\label{Fig2}
Transport properties of a S-Benzene-Te molecule sandwiched between Ni [001] surfaces. In the 
large panels the $I$-$V$ characteristic for both the PA (a) and AA (b). In the left inset 
the MR ratio as a function of bias, and in the right one the transmission coefficient as a
function of energy for the PA and zero bias.}
\end{figure}

In figure \ref{Fig2} we present the $I$-$V$ characteristics for both the PA and AA, the MR
ratio as a function of the bias voltage and the transmission coefficients as a function of energy 
for zero bias in the PA configuration.
From the transmission coefficient it is clear that most of the current through the device is carried
by one broad transmission peak located just below the Fermi level ($E_\mathrm{F}$) for the majority 
spins and just above for the minority. These are associated with a hybrid SS formed from Ni-$d$
orbitals with the $p$-shells of both S and Te. The resulting transport properties are those expected
from the simple discussion of the previous section.

In the PA the two spin-currents are rather similar since no resonant condition is ever
met except for zero bias. Note also that both spin-currents almost saturate at around 
$V=\pm$0.5~Volt, which approximately corresponds to the width of the SS.  Finally it is
clear that the spin polarization of the current, defined as $P=(I^\uparrow-I^\downarrow)/
(I^\uparrow+I^\downarrow)$ ($I^\sigma$ is the spin $\sigma$ current), decreases as the 
bias increases and approximately vanishes for $V\sim$1~Volt.

In the AA the two spin-currents individually present a rather asymmetric bias dependence 
reflecting the two resonant conditions. However the total current is almost completely symmetric
and the tiny asymmetry is associated with the slightly different bonding of S and Te with the
Ni hollow site. Importantly also in this case the current saturates at around $V=\pm$0.5~Volt,
when the resonant condition is not met any longer. This gives a MR ratio with a minimum and negative sign
at around $V=\pm$0.5~Volt. Interestingly the spin-resolved MR ratio $R_\mathrm{MR}^\sigma=
(I_\mathrm{PA}^\sigma-I_\mathrm{AA}^\sigma)/I_\mathrm{AA}$ (note that $R_\mathrm{MR}^\uparrow+
R_\mathrm{MR}^\downarrow=R_\mathrm{MR}$) is rather asymmetric with $V$,
reflecting the asymmetry of the spin-currents.
Note finally that the MR ratio changes sign when the bias is increased
from zero, since we gradually move from the resonant condition for the PA to those for the AA. Similar
results have been recently reported for different anchoring structures \cite{Guo} and
molecules \cite{Kir2}.

%%%%%%%%%%%%%%%%%%%%%%%%%%%%%%%%%%%%%%%%%%%%%
%%%%%%%%%%%%%%%%%%%%%%%%%%%%%%%%%%%%%%%%%%%%%

\section{STM configuration}

\begin{figure}[ht]
\includegraphics[width=0.45\textwidth]{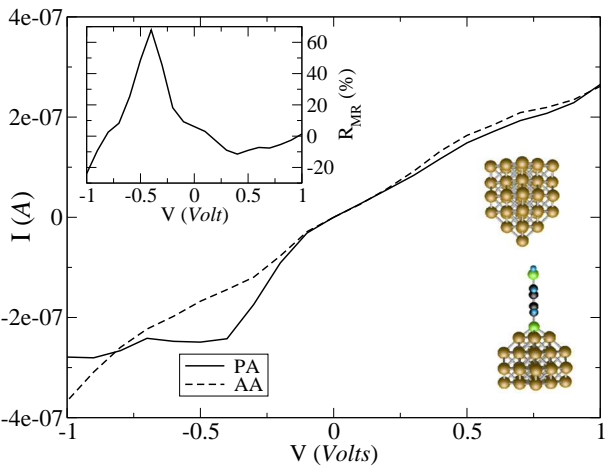}
\caption{\label{Fig3}
$I$-$V$ characteristic for a benzene-thiol-thiolate molecule attached to the hollow site of Ni [001] 
and probed with a Ni STM tip. In the insets the geometrical configuration of the device and
$R_\mathrm{MR}$ as a function of bias.}
\end{figure}

The junction described in the previous section presents rather large currents and the typical 
transmission coefficient approaches $T=1$. This is not the ideal situation for studying resonances,
which in contrast dominate the transport in the case of overall small transmission. For this reason
we move to investigating a STM-type geometry. Here we consider the same benzene molecule attached
to the hollow site of a Ni [001] surface via S and terminated with a thiol group. This latter is
then in loose contact with a single-atom Ni STM tip positioned on top of S (with a Ni-S separation of
4\AA) and shifted perpendicularly to the benzene plane by $\approx$1.2\AA\ (see inset of figure \ref{Fig3}).
In this configuration we probe the $p_z$ orbital of S (i.e. the one perpendicular to the benzene 
plane), which is involved in the transport.

In figure \ref{Fig3} we present both the $I$-$V$ characteristics and the MR for such a device. At variance 
with the previous case the total currents for the PA and AA are rather similar for forward bias, but differ
substantially for the negative voltages. In this case the PA current drastically increases for $V$ between -0.1~Volt
and -0.5~Volt and then saturates. In turns this produces a large positive peak in the MR reaching 
70\% for $V$ around -0.5~Volt. Away from this peak the MR ratio is a smooth function of the bias and
generally it is quite small.

\begin{figure}[ht]
\includegraphics[width=0.45\textwidth]{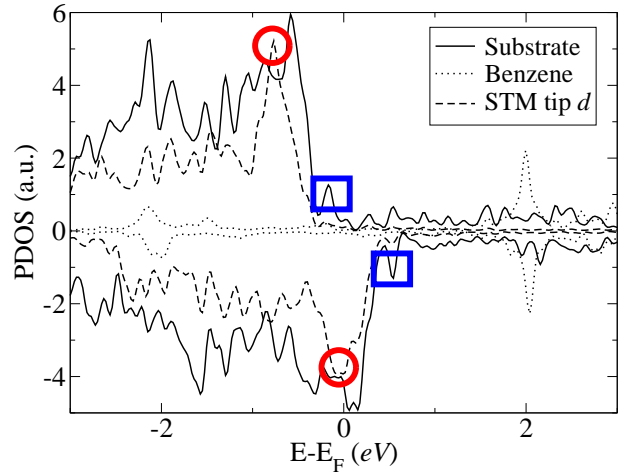}
\caption{\label{Fig4}
Orbital resolved density of state for benzene-thilolate-thiol molecule attached to Ni [001]
and probed by a Ni STM tip. The tip and surface states are indicated respectively with
red circles and blue squares.}
\end{figure}

In order to have a better understanding of this second situation in figure \ref{Fig4} we show the orbital 
resolved density of states for our device, where in particular we project over the Ni [001]
surface, the benzene molecule and the $d$ orbitals of the tip. From the picture we can identify the
two spin-polarized states mainly responsible for the transport. The first is the tip $d$-orbitals band-edge (red circles in figure \ref{Fig4}) 
with majority and minority components approximately 200~meV below the corresponding $d$ 
band-edge of the substrate.
The second is a hybrid SS with spin-polarized components (blue squares) approximately 
200~meV above the corresponding $d$ band-edge of the substrate.

It is then clear that a resonant condition is found whenever the tip $d$ band-edge is aligned 
with SS of the substrate. However, in contrast to the cartoon of figure \ref{Fig1}, the resonant states
of the left-hand side electrode (associated to the Ni [001] substrate) and of right-hand side electrode 
(associated to the tip) are in different positions. The alignment that we obtain from figure 
\ref{Fig4} immediately suggests that, regardless of the magnetization alignment of the electrodes, 
there cannot be any resonant condition for positive bias (current going from the substrate to the tip), 
since the tip state is lower than the SS of the substrate. In contrast a resonant condition can be 
matched at negative bias for PA and both spins.

\begin{figure}[ht]
\includegraphics[width=0.45\textwidth]{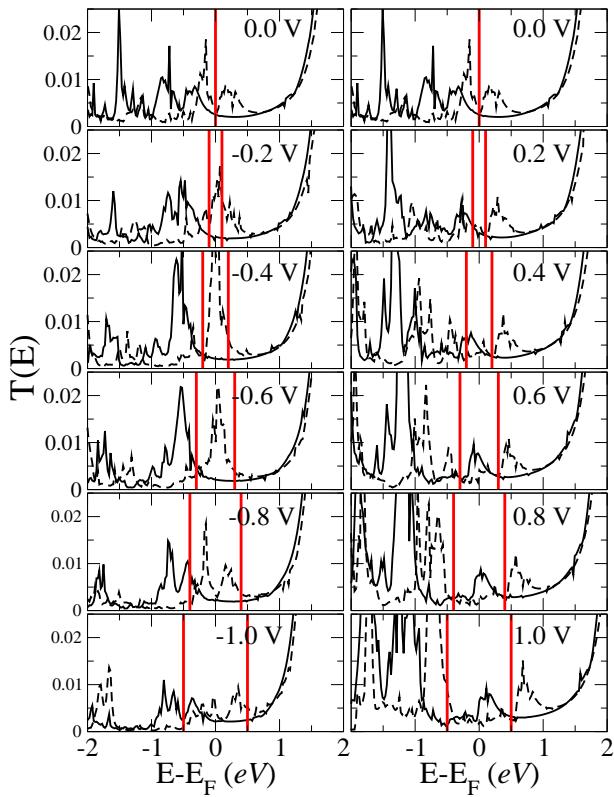}
\caption{\label{Fig5}
Transmission coefficient as a function of energy at different bias for
benzene-thilolate-thiol molecule attached to Ni [001]
and probed by a Ni STM tip: parallel alignment only. 
Note the sharp increase of the transmission for V$\sim$~-0.4~Volt. The solid
(dashed) line represent respectively majority and minority spins. The vertical red lines indicate the
bias window.}
\end{figure}

In figure \ref{Fig5} we demonstrate this resonant condition and present the transmission coefficient
as a function of energy in the PA for various biases. We note that for zero bias $T(E)$ displays two
0.5~eV wide peaks around $E_\mathrm{F}$ for both spin directions. These directly reflect the tip
$d$ orbital band-edge and the spin-polarized SS. As the negative bias increases the tip and substrate
state move to resonance as demonstrated by the large transmission peak at $V\sim$~-0.4~Volt. This
occurs for both majority and minority spins, however only the minority contribute to the current
since only the minority peak falls within the bias window.
Finally as the bias keeps increasing the resonance is destroyed and the enlargement of the
bias window is compensated by a drastic reduction of the transmission peaks. This causes the
current to saturate.

Note that for positive bias no resonant behavior is observed within the bias window at any of the
voltages investigated, although some resonances are observed approximately 1~eV below 
$E_\mathrm{F}$. A similar non-resonant situation is encountered for the AA case and it is not
presented here. 

%%%%%%%%%%%%%%%%%%%%%%%%%%%%%%%%%%%%%%%%%%%%%
%%%%%%%%%%%%%%%%%%%%%%%%%%%%%%%%%%%%%%%%%%%%%

\section{Conclusions}
In conclusion we have investigated resonant transport in organic spin-valves. In the case of
a single resonant state we predict a maximum in the current for the antiparallel alignment 
of the electrodes at a bias comparable with the exchange-splitting of the resonant state.
In contrast in the case of highly asymmetric anchoring structures multiple resonant states
may appear and different resonant conditions can be found. In particular we investigate 
the almost symmetric case of benzene molecule chemically bonded to Ni [001] surface via S and Te, and
that of a STM-type geometry. Importantly the MR ratio is a non-trivial function of bias and can
assume both positive and negative values. Therefore such a strong variation of $R_\mathrm{MR}$
with bias seems to be a distinctive feature of organic spin-valves.

%%%%%%%%%%%%%%%%%%%%%%%%%%%%%%%%%%%%%%%%%%%%%
%%%%%%%%%%%%%%%%%%%%%%%%%%%%%%%%%%%%%%%%%%%%%

\section{Acknowledgments}
This work is supported by Science Foundation of Ireland under the grants SFI02/IN1/I175 and
SFI05/RFP/PHY0062. Computational resources have been provided by the HEA IITAC 
project managed by the Trinity Center for High Performance Computing and by
ICHEC.


\begin{thebibliography}{99}

\bibitem{SS-Review}S.~Sanvito and A.~Reily~Rocha, J. Comput. Theor. Nanosci. {\bf 3}, 624 (2006).

\bibitem{bruce} K.~Tsukagoshi, B.~W. Alphenaar, and H.~Ago, Nature (London) {\bf 401}, 572-574 (1999).

\bibitem{Ralph} J.~R.~Petta, S.~K.~Slater and D.~C.~Ralph,
Phys. Rev. Lett. {\bf 93}, 136601 (2004).

\bibitem{Shi} Z.~H.~Xiong, D.~Wu, Z.~Valy~Vardeny and J.~Shi,
Nature (London) {\bf 427}, 821-824 (2004).

\bibitem{Dediu} V.~Dediu, M.~Murgia, F.C.~Matacotta, C.~Taliani, and S.~Barbanera,
Solid State Commun. {\bf 122}, 181-184 (2002).

\bibitem{Aws3} M.~Ouyang and D.~D. Awschalom, Science {\bf 301}, 1074-1078 (2003).

\bibitem{mmtrans1}H.B.~Heersche, Z.~de~Groot, J.A.~Folk, H.S.J.~van~der~Zant, C.~Romeike,
M.R.~Wegewijs, L.~Zobbi, D.~Barreca, E.~Tondello and A.~Cornia, 
Phys. Rev. Lett. {\bf 96}, 206801 (2006).

\bibitem{mmtrans2}M.-H.~Jo, J.E.~Grose, K.~Baheti, M.M.~Deshmukh, J.J.~Sokol, E.M.~Rumberger,
D.N.~Hendrickson, J.R.~Long, H.~Park and D.C.~Ralph, Nano Lett. {\bf 6}, 2014 (2006).

\bibitem{SOCarbon}
J. Serrano, M. Cardona and T. Ruf, Solid State Commun. {\bf 113}, 411 (2000)

\bibitem{Alq3}S. Pramanik, C-G. Stefanita, S. Bandyopadhyay, N. Harth, K. Garre, M. Cahay,
cond-mat/0508744

\bibitem{Jullier}M.~Julliere, Phys. Lett. A {\bf 50}, 225 (1975).

\bibitem{SSH}A.J.~Heeger, S.~Kivelson, J.R.~Schrieffer and W.P.~Su, 
Rev. Mod. Phys. {\bf 60}, 782 (1988).

\bibitem{Bishop}S.J.~Xie, K.H.~Ahn, D.L.~Smith, A.R.~Bishop, and A.~Saxena,
Phys. Rev. B {\bf 67}  125202 (2003)

\bibitem{roberta}D. Gatteschi, R. Sessoli and J. Villain, {\it Molecular Nanomagnets},
Oxford University Press (Oxford, 2006)

\bibitem{Smeagol}A.R.~Rocha, V.M.~Garc\'{\i}a-Su\'arez, S.W.~Bailey, C.J.~Lambert, J.~Ferrer and 
S.~Sanvito, Phys. Rev. B {\bf 73},  085414 (2006)

\bibitem{Smeagol1}
A.R.~Rocha, V.M.~Garc\'{\i}a-Su\'arez, S.W.~Bailey, C.J.~Lambert, J.~Ferrer and S.~Sanvito, 
Nature Materials {\bf 4}, 335 (2005).

\bibitem{datta}
S.~Datta, {\em Electronic Transport in Mesoscopic Systems}, Cambridge University Press, 
Cambridge, UK, 1995.

\bibitem{Haug}
H.~Haug and A.~P. Jauho, {\em Quantum Kinetics in Transport and Optics of Semiconductors},
Springer, Berlin, 1996.

\bibitem{KS}W.~Kohn and L.J.~Sham, Phys. Rev. {\bf 140}, A1133 (1965).

\bibitem{DFT}P.~Hohenberg and W.~Kohn, Phys. Rev. {\bf 136}, B864 (1964).

\bibitem{siesta}J.~M.~Soler, E.~Artacho, J.~D.~Gale, A.~Garc\'{\i}a, J.~Junquera, P.~Ordej\'on and
D.~Sanchez-Portal, J. Phys. Cond. Matter {\bf 14}, 2745 (2002).

\bibitem{MZeta}O.F.~Sankey and D.J.~Niklewski, Phys. Rev. B {\bf 40}, 3979 (1989).

\bibitem{MZeta1}J.~Junquera, O.~Paz, D.~Sanchez-Portal and E.~Artacho, Phys. Rev. B {\bf 64},
235111 (2001).

\bibitem{MT} N.~Troullier and J.L.~Martins, Phys. Rev. B {\bf43},1993 (1991).

\bibitem{rgf}S.~Sanvito, C.~J.~Lambert, J.~H.~Jefferson, and A.M.~Bratkovsky,
Phys. Rev. B {\bf 59}, 11936-11948 (1999).

\bibitem{usPC}A.R.~Rocha and S. Sanvito, Phys. Rev. B {\bf 70}, 094406 (2004).

\bibitem{Kir1}H.~Dalgleish and G.~Kirczenow, Phys. Rev. B {\bf 73}, 245431 (2006).

\bibitem{Guo}D.~Waldron, P.~Haney, B.~Larade, A.~MacDonald and H.~Guo, 
Phys. Rev. Lett. {\bf 96}, 166804 (2006).

\bibitem{Kir2}H.~Dalgleish and G.~Kirczenow, Phys. Rev. B {\bf 73}, 235436 (2006).











\end{thebibliography}
\end{document}